\documentclass[prl,aps,twocolumn,showpacs,amsmath,amssymb]{revtex4}

\usepackage{graphicx}% Include figure files
\usepackage{dcolumn}% Align table columns on decimal point
\usepackage{bm}% bold math

\def\ee{$(e,e^\prime)$}
\def\eep{$(e,e^\prime p)$}

\def\deep{$^2$H$(e,e^\prime p)$n}
\def\bolddeep{$^{\bm 2}$H${\bm (e,e^\prime p)}$n}
\def\gevsq{(GeV/c)$^2$}

\begin{document}

\title{
%DRAFT III\\ 
The quasielastic \bolddeep{} reaction at high recoil momenta}

%\input author_actual2_pres
%\input author_actual3

%
% apparently revtex4 macros enjoy breaking the affiliations at commas,
% so I trick it by defining \comma
%
\def\comma{,\ }

\author {
P.E.~Ulmer,$^{1}$
K.A.~Aniol,$^{2}$
H.~Arenh\"ovel,$^{3}$
J.-P.~Chen,$^{4}$
E.~Chudakov,$^{4}$
D.~Crovelli,$^{5}$
J.M.~Finn,$^{6}$
K.G.~Fissum,$^{7}$
O.~Gayou,$^{6,8}$
J.~Gomez,$^{4}$
J.-O.~Hansen,$^{4}$
C.W.~de~Jager,$^{4}$
S.~Jeschonnek,$^{4,9}$
M.K.~Jones,$^{1}$
M.~Kuss,$^{4}$
J.J.~LeRose,$^{4}$
M.~Liang,$^{4}$
R.A.~Lindgren,$^{10}$
S.~Malov,$^{5}$
D.~Meekins,$^{11}$
R.~Michaels,$^{4}$
J.~Mitchell,$^{4}$
C.F.~Perdrisat,$^{6}$
V.~Punjabi,$^{12}$
R.~Roch\'e,$^{11}$
F.~Sabatie,$^{1}$
A.~Saha,$^{4}$
R.~Suleiman,$^{13}$
L.~Todor,$^{1}$ and
B.B.~Wojtsekhowski$^{4}$
\vspace{0.1in}
}

\affiliation{
$^{1}$Old Dominion University\comma Norfolk\comma VA 23529\\
$^{2}$California State University at Los Angeles\comma Los
Angeles\comma CA 90032\\
$^{3}$Johannes Gutenberg-Universit\"at\comma D-55099 Mainz\comma Germany\\
$^{4}$Thomas Jefferson National Accelerator Facility\comma Newport
News\comma VA 23606\\
$^{5}$Rutgers\comma The State University of New Jersey\comma
Piscataway\comma NJ 08855\\
$^{6}$College of William and Mary\comma Williamsburg\comma VA 23187\\
$^{7}$University of Lund\comma P.O. Box 118\comma SE-221 00 Lund\comma Sweden\\
$^{8}$Universit\'{e} Blaise Pascal/CNRS-IN2P3\comma F-63177 
Aubi\`{e}re\comma France\\ 
$^{9}$The Ohio State University\comma Lima\comma OH 45804\\
$^{10}$University of Virginia\comma Charlottesville\comma VA 22901\\
$^{11}$Florida State University\comma Tallahassee\comma FL 32306\\
$^{12}$Norfolk State University\comma Norfolk\comma VA 23504\\
$^{13}$Kent State University\comma Kent\comma OH 44242\\
}

\date{\today}

%\twocolumn

\begin{abstract}
The  \deep{}  cross section  was  measured in  Hall  A  of the  Thomas
Jefferson  National   Accelerator  Facility  (JLab)   in  quasielastic
kinematics  ($x=0.96$)   at  a  four-momentum  transfer  squared,
$Q^2$=0.67 \gevsq{}.   The experiment was performed  in fixed electron
kinematics  for recoil  momenta from  zero to  550 MeV/c.   Though the
measured cross section deviates by $1-2\sigma$ from a state-of-the-art
calculation at  low recoil momenta,  it agrees at high  recoil momenta
where final state interactions (FSI) are predicted to be large.
\end{abstract}


%PACS numbers:
%  25.30.Fj: inelastic electron scattering to continuum
%  21.45.+v: Few-body systems
%  21.30.-x: Nuclear forces
%%%%%%%%%  25.10.+s: nuclear reactions involving few-nucleon systems

%\narrowtext

\maketitle

The  deuteron, as  the only  bound two-nucleon  system,  represents the
simplest  manifestation of  the  nuclear force  and  is therefore  the
natural starting point for investigation  of the nature of the nuclear
electromagnetic  current.  The  applicability of  reaction  models for
complex  nuclei  can be  gauged  by the  success  of  these models  in
reproducing  scattering  observables  on  the deuteron.   Further,  by
studying  the deuteron  in extreme  kinematics where  its  short range
structure is emphasized,  one may determine to what extent
its description in  terms of nucleon/meson degrees of  freedom must be
supplemented by inclusion of explicit quark/gluon effects.  This issue is of
fundamental importance to nuclear physics.  Finally, understanding the
deuteron is  required for the  interpretation of a variety of
experiments using the deuteron as an effective neutron target.

Coincidence  \deep{}   reactions  are  particularly   well  suited  to
nucleon-nucleon ({\sl NN}) interaction studies.  Below pion threshold, the
final state is completely  specified.  Further, by judicious choice of
kinematics one can emphasize various aspects of the reaction dynamics
\cite{fabian}.  Accessing the deuteron's short range structure implies
measurements at high recoil momentum, $p_r$ ({\sl i.e.} the momentum of the
undetected recoiling neutron).  However, for certain kinematics,  
final state interactions (FSI) can change the cross
section by an order of  magnitude or more at high $p_r$ (see, {\sl
e.g.}, Fig.~\ref{fig:deep} below).  These large effects result
from  strength at  low initial  proton momentum feeding high $p_r$
due  to $np$ rescattering  in the  final state.  Models which succeed
under such stringent tests may then be applied with some confidence to
infer aspects of the deuteron short distance structure in kinematics
where $p_r$ is large but FSI are minimized.  Such a situation is
expected at large $x$ ($x=Q^2/2m\omega$) in ``parallel kinematics'', 
where the proton is detected along the direction of $\vec q\,$ 
\cite{arenpriv}.

Although there is  a substantial  body of
data   on  this   reaction from facilities other than JLab,  
including   unseparated   cross  sections
\cite{bernheim,turck,breuker,blomqvist} as well  as separations of
                       response                       functions
\cite{vdschaarlt,tamae,vdschaarlandt,frommberger,ducret,bulten,jordan,pellegrino,kasdorp},
various limitations made it difficult to disentangle aspects of the
reaction mechanism from the deuteron short range structure.
For  the cross  section  measurements, limitations  in  the
accelerator energies available at Bates, Saclay, NIKHEF and Mainz
frustrated such attempts  by forcing
measurement of very high recoil  momenta to energy transfers far above
the  quasielastic peak.  Thus,  for the  Turck-Chieze \cite{turck}
and  high $p_r$ Mainz \cite{blomqvist} data, the kinematics were  
in the $\Delta$-region where lack of
knowledge  of  the reaction  mechanism  made  it  difficult to  deduce
aspects of  the deuteron structure.   Although   the  Mainz
measurement sampled  $p_r$  up to  928 MeV/c,  the
kinematics actually  imply that the  bulk of the cross  section arises
from interaction  with the neutron,  leaving the detected proton  as a
spectator.
%%Within this proton  spectator picture, the actual internal
%%momentum probed  in this process is  not the recoil  momentum, but the
%%momentum of  the detected proton  ($\sim 670$ MeV/c).   
Further, since
the  kinematics  were  in  the  $\Delta$-region of  the  inclusive  \ee{}
spectrum, the inclusion of virtual nucleon excitations was required to
obtain agreement with the data.    Although the energy limitation  is not
shared by  SLAC, the maximum current and duty factor  restricted the
range  of   recoil  momenta  there   as  well.  In contrast, JLab's
combination of high beam energy, current and duty factor
allows examining large recoil momenta at or even below quasielastic 
kinematics ({\sl i.e.} $x\ge 1$), making the
extraction  of the  deuteron  structure less  model-dependent.

The  \deep{}  separation  experiments  have also  been  restricted  in
kinematical coverage for the  reasons stated above.  Nonetheless, they
have revealed  gaps in  our understanding.  Various  calculations have
difficulty       reproducing      both      $R_L$       and      $R_T$
\cite{vdschaarlandt,ducret,jordan}.  The $R_{LT}$ response and related
$A_{\phi}$    asymmetry    \cite{vdschaarlt,frommberger,ducret,bulten}
indicate the need for relativistic treatments but problems still exist
in reproducing  the data.   For momentum transfers  up to  $Q^2\sim 1$
(GeV/c)$^2$ a nonrelativistic  calculation supplemented with the most
important leading  order relativistic contributions has  been shown to
agree quite  well with a covariant approach  \cite{beck}.  More recent
calculations  of  Jeschonnek,  Donnelly  and Van  Orden  \cite{sabine}
suggest  that the  bulk  of the  relativistic  effects may  be in  the
nucleon current  operator, as opposed  to the nuclear  dynamics.  They
obtained  good agreement  between a  manifestly  covariant calculation
involving the  Gross equation \cite{gross}  and a calculation  using a
nonrelativistic  wave  function  and  a relativistic  nucleon  current
operator.   In contrast,  a  calculation with  a nonrelativistic  wave
function  and current  operator  drastically failed  to reproduce  the
results  of the covariant  calculation.  These  calculations highlight
the  importance of  incorporating relativity  properly.  Also,  if the
relativistic effects are  mostly in the current operator,  they can be
incorporated for heavier  nuclei using the same approach.   Due to its
more tractable nature, the deuteron provides a testing ground for such
an approach.

Our experiment (Experiment E94-004) consisted of measuring the unseparated
cross section  in quasielastic kinematics  out to high $p_r$.   It was
performed in  Hall A  of JLab using  the high  resolution spectrometer
pair.   A  detailed  description  of  the Hall  A  instrumentation  is
currently being  prepared \cite{nim}.  Electrons  from the accelerator
were  incident on  a 15  cm long  liquid deuterium  target.  Scattered
electrons and knockout protons were detected in the ``left'' (formerly
``electron'')   spectrometer  and   ``right''   (formerly  ``hadron'')
spectrometer, respectively, each with nominally 6.0 msr collimators at
their front  ends.  Each spectrometer  was equipped with  its standard
detector   package  consisting   of  a   pair  of   VDC's   for  track
reconstruction and  a scintillator  array for trigger  definition.  In
addition, the  electron spectrometer included  an atmospheric pressure
CO$_2$ threshold \v Cerenkov detector for $\pi^-/e$ discrimination.

Kinematics are
given in Table~\ref{tab:kin}.  The beam energy and electron
spectrometer angle and central momentum were kept fixed  throughout the
experiment  at values  corresponding to  the top  of  the quasielastic
peak. The  recoil momentum was  varied by  changing the  proton spectrometer
angle starting  from parallel  kinematics ({\sl i.e.}  protons detected
along the  three-momentum transfer, $\vec q\,$) where  $p_r=0$ to more
backward  angles resulting  in  a maximum  central  value of  $p_r=500$
MeV/c.   The  proton  spectrometer  momentum  setting was  varied  in
conjunction with the  angle so that the central
kinematics were fixed near the deuteron  breakup energy of 2.2 MeV.  
Obstruction from one of the scattering chamber support
posts resulted in a fairly wide spacing between the
last two settings and a  gap in the  measured spectrum {\sl
vs.} $p_r$.

\begin{table}
\caption{\label{tab:kin}  Kinematics (central  values).   The electron
kinematics  were fixed  throughout  the experiment  with incident  and
scattered  energies  of  $E=3.110$  GeV  and  $E^\prime=2.742$  GeV,
respectively, and scattering angle of $\theta_e=16.06^\circ$, leading
to central values of the four-momentum transfer squared of $Q^2=0.665$
\gevsq{}  and $x=Q^2/2m\omega=0.964$ (where  $\omega$ is  the electron
energy transfer  and $m$  is the proton  mass).  Listed below  are the
recoil  momentum ($p_r$), the  proton momentum  ($p$), and  the proton
angle ($\theta_p$) relative to the beam direction.  All quantities are
expressed  in  the laboratory  system.   Also  shown  are the  average
luminosities corrected  for estimated  beam  related target  density
changes.}
\begin{ruledtabular}
\begin{tabular}{cccc}
$p_r$  & $p$    & $\theta_p$ & ${\cal L}_{avg}$            \\
MeV/c  & GeV/c  &    deg     & cm$^{-2}\,\cdot\,$s$^{-1}$  \\ \hline
   0   & 0.885  & $-$58.76   & 6.50E+37               \\
 100   & 0.874  & $-$65.22   & 2.16E+38               \\
 150   & 0.864  & $-$68.47   & 3.48E+38               \\
 200   & 0.851  & $-$71.73   & 3.59E+38               \\
 275   & 0.823  & $-$76.72   & 3.71E+38               \\
 300   & 0.812  & $-$78.41   & 3.76E+38               \\
 500   & 0.689  & $-$92.78   & 3.67E+38               \\
\end{tabular}
\end{ruledtabular}
\end{table}

Measurement of the elastic $^1$H\eep{} reaction, taken with the
spectrometers at the \deep{} $p_r=0$ setting, served as a
normalization check.  The measured yield was compared to a simulation
\cite{mceep}
using the Simon $et$ $al.$ \cite{simon} parameterization for 
$G_{M_p}$ and the $G_{E_p}/G_{M_p}$
ratio measured by Jones $et$ $al.$ \cite{jones}.  The simulation
included acceptance averaging as well as radiative folding
\cite{motsai}.  In order
to remove the ill-defined acceptance edges a technique involving
``$R$-functions'' was used \cite{r-functions}.  Through the $R$-functions, a
multi-dimensional contour in the space of the target variables
and equidistant from a pre-defined boundary was
defined and all events outside the contour were rejected.  
The same contour was used in the simulations and for the $^1$H\eep{}
and \deep{} data.  The ratio of integrated yield for the
simulation to that for the $^1$H\eep{} data was 1.054, amounting to a 5.4\%
correction for the \deep{} data.

For the \deep{} data, beam currents ranged from 10 $\mu$A to
100 $\mu$A.  Since the electron kinematics were fixed, the electron
arm served as a measure of the product of electronic deadtime and
target density; the electronic deadtime for the hadron arm was assumed
to be negligible, since its trigger rate never exceeded 10 kHz.
Computer deadtime was determined from the ratio of coincidence raw
triggers to recorded events.  In order to improve the
reals-to-accidentals ratio, especially important at high $p_r$,
a consistent vertex from both spectrometers was required and a 
cut on the missing mass was imposed.  Finally, a cut on the summed
analog signal from the \v Cerenkov
detector was used to reduce the already small
contamination from $\pi^-$ events in the electron arm. 

The aperture/magnetic model of each spectrometer was tested by
measuring a series of ``white'' spectra, scanned in overlapping
momentum steps.  The relative spectrometer acceptance was
then extracted by an iterative procedure.  For the $R$-function cuts
described above, the results were in excellent agreement 
with the simulated phase space.

%As a test of the aperture/magnetic model of the spectrometers, a
%series of ``white'' spectra were scanned in overlapping steps 
%across the focal planes of
%each spectrometer.  This was done by making small variations of the
%magnetic fields and acquiring prescaled, single-arm data for each
%spectrometer.   An iterative procedure was used to determine the shape
%of the cross section as well as the relative spectrometer 
%acceptance as a function of the
%dispersion, $\delta$ (defined as the percentage deviation of a given
%particle's momentum from the spectrometer central momentum value).
%This relative acceptance was compared to the simulated phase space.
%Both data and simulation were obtained for
%the $R$-function cuts described above.  The results, after applying
%overall vertical scale factors, are shown in 
%Fig.~\ref{fig:effic}.  Within the vertical bars, defined by
%$\delta=\pm3.5$\%, the agreement is essentially perfect.   The
%\deep{} data were also subjected to these $\delta$ cuts, as was the
%$^1$H\eep{} calibration described above.

%\begin{figure}
%\includegraphics{figs/effic_fig.eps}
%\caption{\label{fig:effic} The relative acceptance {\sl vs.} 
%$\delta$ extracted from
%the data compared to the simulation.  The top panel is for the electron
%arm and the bottom is for the proton arm.  The vertical lines at
%$\delta=\pm 0.035$ indicate the cuts which were applied to both data
%and simulations in extracting the cross sections.  The differing
%slopes of the two distributions results from the difference in
%spectrometer angles and consequent extended target viewing angles.}
%\end{figure}

The total systematic uncertainty in the cross sections was estimated to
be roughly  8\%, nearly independent  of $p_r$, except near  zero where
it grows sharply due to the shrinking of the phase space volume.  The
kinematic related uncertainties were  estimated by computing the cross
section at  the center of  a given bin  in $p_r$, further  weighted by
$p_r^2$ to roughly account for the phase space volume, and then making
variations of  each kinematic quantity  in turn.  The  quantities were
varied in a correlated fashion, constrained by the elastic $^1$H\eep{}
kinematics and by independent measurement of the beam energy, obtained
by measuring the position of the beam at a point of high dispersion in
an eight  dipole arc  section.  The quadrature  sum of  all kinematics
related uncertainties was  less than 2.5\% for all  recoil momenta not
too close  to zero.  Other  uncertainties were estimated to  be: 2.0\%
(electronic  deadtime $\times$  target density),  2.0\%  (electron arm
solid angle; not accounted  for in the $^1$H\eep{} normalization where the
acceptance was limited by the proton arm), 1.0\% (beam charge relative
to   normalization),   3.0\%    (radiative   correction)   and   5.8\%
(normalization, consisting  of 3.6\%  from uncertainty in  the elastic
form factors,  2.0\% uncertainty from kinematics and  4.1\% from other
uncertainties  specific  to  the normalization  measurement).   Adding
these uncertainties quadratically yields a  value of 7.2\% taken to be
independent of $p_r$.

The results for data along with various calculations 
are shown in Fig.~\ref{fig:deep}.  The top panel shows the radiatively
corrected ``reduced'' cross section:
\[ \sigma_{\rm red}\equiv \frac{d^5 \sigma}{d\Omega_e d\omega d\Omega_p}
\times \frac{1}{f_{\rm rec} K \sigma_{\rm CC1}} \]
where $K$ is a kinematic factor, $\sigma_{\rm CC1}$ is the
half-off-shell electron proton cross section of de Forest
\cite{deforest} and $f_{\rm rec}$ is a recoil factor which arises from
the integration over missing mass.  This division removes most of the kinematic
dependence, except for the $p_r$ dependence, and results in a smooth
spectrum {\sl vs.} $p_r$.   The bottom panel shows the relative
deviation of data and theoretical predictions from the ``full'' calculation 
of Arenh\"ovel \cite{arenhovel} (described below).  Also
shown in the bottom panel is the systematic error band, arbitrarily
placed vertically.  The data were radiatively corrected by multiplying
the measured cross section in each $p_r$ bin by the ratio of yield 
without and with radiative effects (internal and external),
both estimated using Arenh\"ovel's full calculation.
The calculation was radiatively folded using the model of Borie and
Drechsel \cite{borie}.  Both data and simulations were cut on
the missing mass from $-$3.5 MeV to 10.5 MeV, resulting in a
weak dependence on $p_r$ for the radiative correction factor.
A straight-line fit to this dependence was used.
%%Without such a cut there would be strong feeding
%%from low to high $p_r$ caused by radiation of hard photons and a
%%consequent strong dependence of the correction factor on $p_r$.

\begin{figure}
\includegraphics{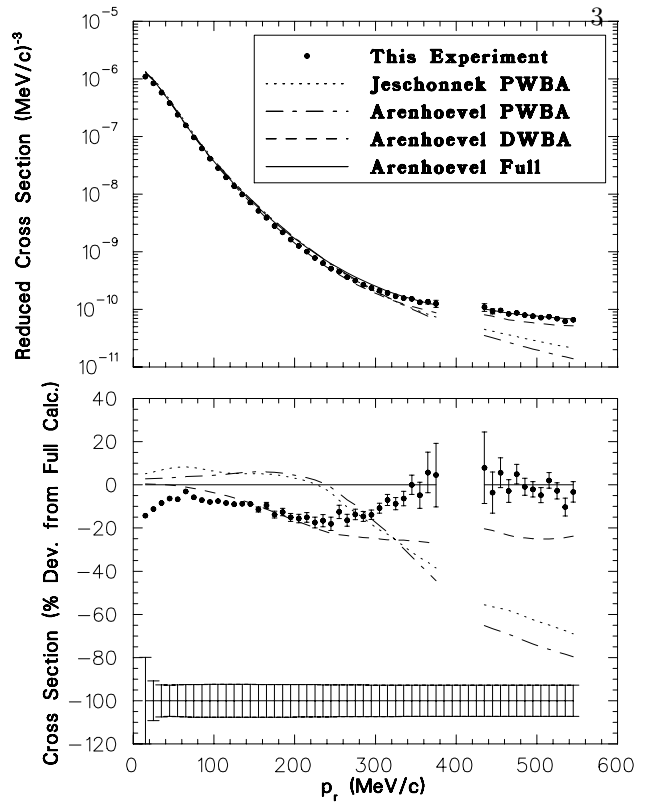}
\caption{\label{fig:deep} Top panel:  The reduced \deep{} cross section 
for this experiment along with various model
calculations (see the text for details).  Bottom panel: Cross sections 
for data and calculations 
shown as percentage deviations from the ``full'' calculation of
Arenh\"ovel.  Also shown is the systematic error band ($\pm 1
\sigma$), arbitrarily placed on the vertical axis.  This error band 
contains an overall 7.2\% contribution added in quadrature with the
kinematic contribution, the latter varying slightly with $p_r$.}
\end{figure}

All of  the models were acceptance averaged  using MCEEP \cite{mceep};
the  calculations of  Arenh\"ovel were  performed on  a grid  over the
experimental acceptance  and interpolated for each  event, whereas the
calculations  of Jeschonnek  \cite{sabine} were  incorporated directly
into the Monte Carlo simulation  program as a subroutine package.  The
Jeschonnek calculation was in the plane wave Born approximation (PWBA)
and used a fully  relativistic single-nucleon current operator with an
alternate three-pole parameterization of  the MMD nucleon form factors
\cite{mmd}  and the Argonne  V18 two-body  interaction \cite{argonne}.
The calculations of Arenh\"ovel included relativistic contributions of
leading order in $p/m$ to the kinematic wave function boost and to the
nucleon current.   The Bonn  $r$-space {\sl NN}  potential \cite{bonn}
and dipole nucleon form factors were used.  The various curves are for
PWBA, distorted  wave Born  approximation (DWBA, which  includes FSI),
and  the  ``full''   calculation  which  also  includes  non-nucleonic
currents: meson exchange currents and virtual nucleonic excitations.

The  two PWBA  calculations are  reasonably close  to each  other, but
deviate  at  high  $p_r$,  presumably  largely due  to  the  different
{\sl NN} potentials employed.  For $p_r>300$ MeV/c the PWBA
fails  completely,  whereas the  full  calculation  including FSI  and
significant  contributions  from   non-nucleonic  degrees  of  freedom
results in satisfactory agreement  with the data.  For $p_r<100$ MeV/c
the full theory deviates from the data by roughly $1\sigma$ growing to
about  $2\sigma$ at  $p_r\sim 200-300$  MeV/c, adding  statistical and
systematic  errors  quadratically.  It  is  important  to resolve  any
possible discrepancies  in this low $p_r$ region,  especially in light
of  some of  the neutron  form factor  measurements which  exploit the
deuteron  in  this  kinematic  region.   Clearly,  additional,  higher
precision measurements would be  helpful in clarifying this situation.
An  experiment to  systematically  study this  reaction  over a  broad
kinematical range has already been proposed and conditionally approved
by the  JLab Program  Advisory Committee \cite{newprop}.   This future
experiment promises significantly  smaller systematic errors, based on
recently acquired  experience with  the experimental apparatus  and on
inclusion  of   a  larger  set   of  experimental  cross   checks  and
calibrations.

In summary, we have measured the \deep{} cross section at
$Q^2=0.67$ \gevsq{} and $x=0.96$ for recoil momenta up to 550 MeV/c.
The data indicate large FSI and substantial non-nucleonic effects at
high $p_r$.  The full calculation overestimates the cross section by
$1-2\sigma$ at small recoil momenta.

We acknowledge the outstanding support of the staff of the 
Accelerator and Physics
Divisions at Jefferson Laboratory that made this experiment successful.
This work was supported in part by the U.S. Department of
Energy  Contract No.  DE-AC05-84ER40150 under  which  the Southeastern
Universities Research Association (SURA) operates the Thomas Jefferson
National Accelerator  Facility, other Department  of Energy contracts,
the  U.S.  National  Science Foundation,  the French  Commissariat \`a
l'Energie Atomique  and Centre  National de la  Recherche Scientifique
(CNRS) and the Deutsche Forschungsgemeinschaft (SFB 443).

\end{document}